\documentclass[twocolumn,preprintnumbers,amsmath,amssymb]{revtex4}
\usepackage{graphicx}
\usepackage{dcolumn}
\usepackage{bm}

\begin{document}
\title{Interlayer repulsion and decoupling effects in 
stacked turbostratic graphene flakes}
\author{Julia Berashevich and Tapash Chakraborty}
\affiliation{Department of Physics and Astronomy, The University
of Manitoba, Winnipeg, Canada, R3T 2N2}

\begin{abstract}
The behavior of stacked graphene flakes is found to be governed by the 
strength of the repulsive interactions that arise due to the orthogonality 
of interlayer $\pi$ orbitals. Therefore, the decoupling effect in AA 
stacked layers is a result of the repulsion being dominant over the 
orbital interactions while misorientation of 2$^{\circ}$-5$^{\circ}$ 
is an attempt by the system to suppress that repulsion. For misorientated 
graphene, in the regions of superposed lattices in the Moir\'e pattern, the
repulsion between the layers manifest itself as lattice distortion by forming a bump.
\end{abstract}

\maketitle
Ever since the discovery of graphene obtained by mechanical exfoliation of graphite
\cite{nov}, the enormous interest on this material that exhibits many unique
electronic properties is yet to subside \cite{review}. However, the focus has
recently shifted to the epitaxial growth of graphene \cite{varchon,naturep,hass}
as being a more advanced process to control the graphene geometry that would be
beneficial for its application in electronics. Here, the multilayer structures
are of particular interest due to their high robustness and because their electronic
properties can be manipulated through variation of the number and orientation of
the layers \cite{hass,naturep,varchon}. A clear understanding of the influence of
the layer orientation on the electronic properties of graphene is one of the most
urgent issues that need to be resolved. It was observed by several groups that the
AA stacked epitaxial systems experience a misorientation of $\sim$2$^{\circ}$-5$^{\circ}$
\cite{hass,naturep,varchon} between the layers and exhibit the electronic properties
of a monolayer graphene because of decoupling of the layers, which also occurs in
graphene obtained by chemical vapor deposition \cite{reina} and ultrasonicating
graphite \cite{warner}. The decoupling is also observed for another commensurate
angle near 30$^{\circ}$ \cite{hass}. This behavior is different from that known for
the AB stacked bilayer graphene \cite{mccann,latil}. Several theoretical works
were reported on the electronic properties of twisted graphene \cite{latil,shall,lopes}
and indeed the monolayer behavior for $\theta\rightarrow 0^{\circ}$ has been confirmed.
However, no proper explanation of the underlying physical reasons for the electronic
decoupling is available as yet.  

The carbon atoms in graphene are connected by the covalent $\sigma$ bonds in $sp^2$ 
hybridization. After forming the hexagonal lattice, each carbon atom has one extra 
valence electron residing on the $\pi$ orbital. Those $\pi$- electrons generate
the $\pi$ and $\pi^{\ast}$ bands which meet and display linear dispersion in the 
vicinity of the K-points \cite{review}. As the $\pi$ orbitals are orthogonal to the 
graphene plane, in a system of two stacked layers the $\pi$ orbitals interact. When 
the lattices of the two layers are superposed, the conditions for a perfect overlap 
of the $\pi$ orbitals are created. However, decoupling between the layers as seen 
in the experiments requires an opposite behavior which inspired us to investigate 
this phenomena in the present work with the help of the quantum chemistry methods 
\cite{nbo}. In graphene the $\pi$ electrons completely fill the bonding orbitals 
thereby generating a closed electron shell. When two closed-shell systems are stacked, 
the interlayer interaction of their $\pi$ orbitals is repulsive. The bonding and 
antibonding interactions which may be different from zero individually, will overall 
cancel each other and the $\pi$ electrons are expelled from the overlap 
region. 

\begin{figure}
\includegraphics[scale=0.40]{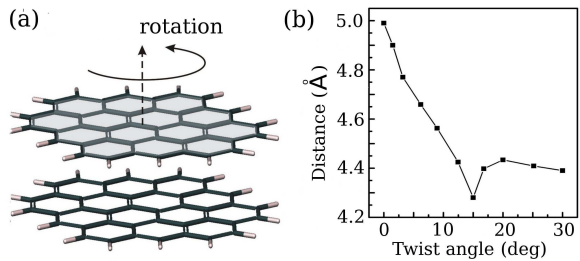}
\caption{\label{fig:fig2} a) The system of graphene flakes stacked at the twist angle 
$\theta$=0$^{\circ}$ and interlayer distance 3.34 \AA. b) Dependence of the 
equilibrium interlayer distance on the twist angle.}
\end{figure}

The interlayer coupling is the parameter most widely used in the tight-binding 
approximation, which is why we intend to investigate how the coupling is influenced 
by the repulsion. We considered two graphene flakes (Fig.~\ref{fig:fig2} (a)) with 
non-linear edges terminated by hydrogen atoms thereby excluding the appearance of 
localized states which would closely resemble the infinite case.
To estimate the interlayer coupling we employ the natural bond orbital (NBO) analysis 
\cite{nbo}. This method uses the electron density distribution obtained after the DFT 
calculation (with the UB3LYP/6-31+G* hybrid exchange-correlation functional) to build the 
natural bond orbitals $\Omega$. The interaction matrix $F_{i,j}$ required to perturb 
the natural bond orbitals $i$ and $j$ into the molecular orbitals is then created for 
all the double bonds involved in the interaction between two molecular fragments. Each 
element $F_{i,j}$ is for the electronic coupling between the bonds, while the sum of 
$F_{i,j}$ defines the interlayer coupling between the two flakes. Within the NBO 
procedure the Pauli exclusion principle is applied to the outer and inner nodes thereby 
preserving the interatomic orthogonality and its for electrons on the same orbital. That 
is an important advantage of using the NBO for closed-shell systems like graphene. 
The occupancy charge transfer from one flake to another is then calculated as the 
sum of $\Omega_i\rightarrow\Omega^{\ast}_j$ transfers between the donor $\Omega_i$ 
and acceptor $\Omega^{\ast}_j$ orbitals,
$Q_{1\rightarrow2}=\sum_{i,j}q_i F_{i,j}^2/(\epsilon_i-\epsilon_j)^2,$
where $\epsilon_i,\epsilon_j$ are the orbital energies. The transfer is calculated for 
the stabilizing interactions, i.e. when the second-order interaction 
energy $-2F_{i,j}^2/(\epsilon_i-\epsilon_j)$ is positive.

The orthogonality of the $\pi$ orbitals when the two layers are superposed can be 
destroyed by a rotation of the flake as shown in Fig.~\ref{fig:fig2} (a). Increasing 
the twist angle from $\theta=0^{\circ}$ to 30$^{\circ}$ we calculate the occupancy 
transfers between the layers $Q_{1\rightarrow2}$ ($Q_{2\rightarrow1}$) and the 
off-diagonal element $F_{i,j}$ for an interlayer distance of 3.34 \AA\ (typical 
for graphite). The results are collected in Table~\ref{tab:table1}. To estimate the 
interlayer coupling we took an average magnitude of the interaction matrix $\langle 
F_{1-2} \rangle$ defined as $\sum_{i,j}F_{i,j}$ for every interaction between the 
$\pi$ bonds and divided by the number of double bonds in one layer.

\begin{table}
\caption{\label{tab:table1} Dependence of the electronic properties on the twist 
angle $\theta$. $Q_{1\rightarrow 2 (2\rightarrow 1)}$ is the interlayer occupancy 
charge transfer. $\langle F_{1\rightarrow2}\rangle$=$\sum_{i,j}F_{i,j}$/21 is the 
average value for the off-diagonal element. $E_{H}$ and $\Delta E_{H}$ are the HOMO 
orbital energy and (HOMO-(HOMO-1)) splitting, respectively. $E_{G}$ is the HOMO-LUMO 
gap. $\Delta E_{tot}$ is the difference in the total energy between the current 
conformation and that for $\theta=0^{\circ}$ (numbers in brackets are given for the
MP2 calculations \cite{nbo}). AB is a Bernal stacking. All values (except for $Q$)
are in eV.}
\begin{ruledtabular}
\begin{tabular}{c|c|c|c|c|c}
$\theta$, $^{\circ}$ & $Q_{1\rightarrow2}$ ($Q_{2\rightarrow1}$), \={e} & 
$\langle F_{1\rightarrow2}\rangle$& $E_{H}(\Delta E_{H})$& 
$E_{G}$& $\Delta E_{tot}$\\\hline
0.0  & 0.0018 (0.0019)& 0.03 & -3.94 (0.835) & 1.32 & 0.0 (0.0)\\
1.5  & 0.0029 (0.0027)& 0.04 & -3.95 (0.812) & 1.33 & -0.02 (-0.02)\\
3.0  & 0.0038 (0.0035)& 0.05 & -3.96 (0.765) & 1.35 & -0.06 (-0.06)\\
6.0  & 0.0064 (0.0057)& 0.11 & -4.02 (0.638) & 1.45 & -0.19 (-0.21)\\
9.0  & 0.0110 (0.0098)& 0.18 & -4.09 (0.476) & 1.60 & -0.36 (-0.43)\\
12.0 & 0.0152 (0.0129)& 0.22 & -4.19 (0.294) & 1.79 & -0.46 (-0.65)\\
15.0 & 0.0171 (0.0157)& 0.24 & -4.34 (0.012) & 1.99 & -0.63 (-0.83)\\
18.0 & 0.0184 (0.0156)& 0.24 & -4.31 (0.077) & 1.92 & -0.66 (-0.93)\\
21.0 & 0.0179 (0.0159)& 0.25 & -4.24 (0.245) & 1.78 & -0.70 (-0.99)\\
26.0 & 0.0141 (0.0136)& 0.23 & -4.15 (0.459) & 1.64 & -0.67 (-0.96)\\
30.0 & 0.0140 (0.0148)& 0.21 & -4.13 (0.510) & 1.65 & -0.64 (-0.88)\\
AB   & 0.0177 (0.0181)& 0.24 & -4.35 (0.108) & 2.01 & -0.86 (-1.31)\\
\end{tabular}
\end{ruledtabular}
\end{table}

For the 0$^{\circ}$ twist angle the extremely weak $\pi$ interlayer interaction 
occurs mainly between the double bonds located on top of each other such that 
the stacked $\pi$ orbitals are orthogonal. These interactions are found to cancel 
when the graphene lattice is relaxed within the graphene plane. The interaction 
for other combination of double carbon bonds (the next-nearest neighbor interlayer 
interactions) is also almost zero because of their remoteness from each other. 
Therefore for $\theta=0^{\circ}$ the interlayer coupling and the occupancy transfer 
are close to zero ($Q_{1\rightarrow2}$ is compensated by $Q_{2\rightarrow1}$ 
within the intermolecular threshold of 0.0013 eV \cite{nbo}). 

The flake rotation leads to the suppression of repulsion between the layers and 
a manifestation of the next-nearest neighbor interlayer interactions thus enhancing 
the interlayer coupling. The effect of vanishing repulsion is reflected by an
increase of the electron density in the overlap region (the forces expelling the 
$\pi$ electrons from the overlap region are reduced). This leads to an
enhancement of the occupancy transfer $Q_{1\rightarrow2(2\rightarrow1)}$
along with the interaction matrix $F_{i,j}$. Thus,
for those $\pi$ orbitals that are orthogonal at $\theta=0^{\circ}$, 
the $F_{i,j}$ increases from 0.22 eV (at $\theta=0^{\circ}$) to $\sim$ 0.25 eV. For the Bernal 
stacking (AB) the repulsion between layers reaches a minimum and together with 
the high orbital overlap (many of the double bonds belonging to the different 
layers are aligned thus leading to a large overlap) one generates the highest occupancy 
charge transfer between the graphene layers (AB in Table~\ref{tab:table1}). 
Therefore, for AB stacking the value of $F_{i,j}$ is increased up to 0.30 eV for 
the nearest interlayer interactions, while for the next nearest neighbor interactions 
it varies in the range of 0.08-0.16 eV. These results are consistent
with the experimental data for AB stacking \cite{malard}. Because with rotation 
an increase of $F_{i,j}$ is not significant for those $\pi$ orbitals which have been orthogonal 
at $\theta\rightarrow 0^{\circ}$, an enhancement of the occupancy transfer occurs mostly
due to the development of the next-nearest neighbor interlayer interactions. 

Therefore, for $\theta\approx15^{\circ}$ an increase in the occupancy transfer 
reaches its maximum as the orbital overlap is still significant but the repulsion 
is already weak (interlayer location of the double bonds is somewhat parallel 
thereby providing the efficient overlap). Moreover, at this angle one layer clearly 
becomes a donor while the other is an acceptor such that $Q_{1\rightarrow2}>
Q_{2\rightarrow1}$. For $\theta > 15^{\circ}$ the charge transfer starts to 
diminish because the interacting $\pi$ orbitals are becoming distant and misoriented 
from being parallel thereby suppressing their overlap. Misorientation is large for both 
the nearest and next-nearest neighbor interactions, and at $\theta\approx30^{\circ}$ 
the misorientation reaches its maximum. Our results also reveal that for $\theta\approx
30^{\circ}$ the significant misorientation of the interlayer separated $\pi$ orbitals 
would induce a relocation of the double bonds to new positions such that the occupancy 
transfer not only diminishes but also becomes equal in both directions $Q_{1\rightarrow2}
\approx Q_{2\rightarrow1}$. Therefore, the twist angles $n\pi$/6 ($n$ is the positive 
integral number) are the critical points for which the reduction of interlayer coupling 
is induced by the suppression of the $\pi$ orbital overlap and by initiation of 
relocation of the double bonds. The twist angles $n\pi$ are another specific points 
for which the interlayer coupling is suppressed to almost zero because of the 
significant repulsion between the layers as the interlayer $\pi$ orbitals become 
again orthogonal. The same effect occurs for the angles 2$n\pi$/3 if 
the rotation axis is located at the center of the carbon ring.

If the conformation changes are not allowed in the system where the repulsive 
interactions dominate, an increase in energy for pairs of interacting $\pi$ electrons 
occurs and one should consider the alteration of their orbital energies for both 
occupied and unoccupied orbitals (HOMO and LUMO) and the total energy 
($E_{tot}$). The important proof of the presence of a repulsion between the stacked 
flakes is the appearance of destabilization as the flakes move closer because then
the repulsion is significantly enhanced. Thus, with a reduction of the interlayer 
distance from 3.34 \AA\ to 2.5 \AA\ the total energy is enhanced by 22.35 eV. Moreover, 
this leads to a reduction of $F_{i,j}$ by half for orthogonal interactions.
The total energy additionally deviates with layer rotation being a function of the 
orbital mixing altered by the rotation (see $\Delta E_{tot}$ in Table~\ref{tab:table1}). 
To account for the contribution of the van der Waals interactions we applied the local 
M\o ller-Plesset second order perturbation theory (local MP2/6-31+G*)
which provided some reduction in the $\Delta E_{tot}$.
 
The interaction of the stacked graphene layers initiates the perturbation of their 
$\pi$ orbitals thereby modifying the orbital energies that are identical in the two 
layers before perturbation. This shift can be estimated as 
$$E_{1(2)}\approx e_{0}\mp H_{12}\pm (e_{0}\mp H_{12})S_{12},$$
where $e_{0}$ and $E_{1,2}$ are the molecular $\pi$-orbital energies before (for 
identical graphene flakes $e_{1}=e_{2}=e_{0}$) and after perturbation, respectively. 
$S_{12}$ is the orbital overlap. $H_{12}$ is the intrinsic interaction integral 
considering the contribution from the electron-electron interactions and particularly 
of its repulsive part to the orbital energies $E_{1(2)}$. The orbital energies 
and splitting of HOMO orbitals with rotation are presented in the Table~\ref{tab:table1}. 

\begin{figure}
\includegraphics[scale=0.42]{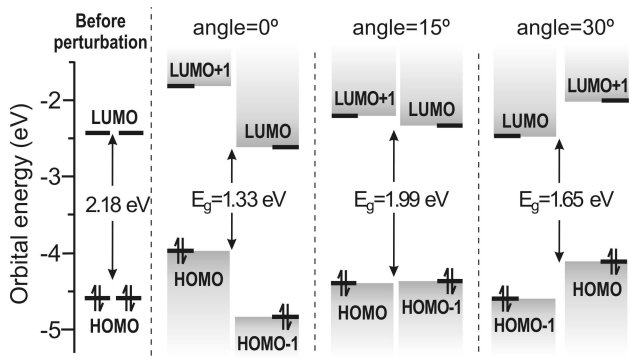}
\caption{\label{fig:fig3} Energetics of the HOMO and LUMO orbitals after the
application of interlayer interactions between the flakes.}
\end{figure}

In Fig.~\ref{fig:fig3} we show the energy diagram for alteration of the HOMO 
and LUMO orbitals due to stacking of graphene layers with the twist angles 
0$^\circ$, 15$^\circ$ and 30$^\circ$, which we argued above as points of particular 
interest. For a graphene flake of such a small size (11.4$\times$9.8 \AA) the 
HOMO-LUMO gap is quite large that is due to the strong confinement effect 
\cite{son-ritter}, while hydrogenation of the edges is a factor enlarging the 
gap even further \cite{barone}. For zero twist angle the interlayer located $\pi$ 
orbitals are perfectly orthogonal. However, the $\pi$ valence electrons 
are expelled from the overlap region by the repulsive forces that suggest
their zero overlap ($S_{12}$=0). Therefore, contribution of 
the $(e_{0}\pm H_{12})S_{12}$ term into the orbital 
energies $E_{1(2)}$ vanishes. This suggests almost no $\pi$ orbital mixing, so 
that the splitting of two HOMO orbitals ($E_{1}-E_{2}$) is defined by the repulsive 
part of the electron-electron interactions ($H_{12}$) and reaches its maximum 
($\Delta E_{H}$=0.835 eV). Our simulation results shown in Fig.~\ref{fig:fig3} 
reveal that for $\theta=0^{\circ}$ the large splitting of the (HOMO-(HOMO-1)) 
and ((LUMO+1)-LUMO) initiates the moving of HOMO and LUMO orbitals closer to each 
other thereby reducing the HOMO-LUMO gap to $E_G$=1.33 eV. However, the gap 
attributed to each layer HOMO-(LUMO+1) and (HOMO-1)-LUMO is being similar to that 
before perturbation.

The rotation of the flakes breaks the orthogonality of $\pi$ orbitals thereby 
inducing the non-zero overlap $S_{12}$ such that the contribution of the $(e_{0}\pm 
H_{12})S_{12}$ term to $E_{1(2)}$ orbital energies is non-zero. For the 
twist angle $\theta\approx15^\circ$ the combination of a large orbital overlap 
and low repulsion leads to almost zero splitting of the (HOMO-(HOMO-1)) thereby 
increasing the HOMO-LUMO gap to 1.99 eV. For the twist 
angle $\theta=30^\circ$ the interlayer $\pi$ orbitals become significantly 
distant and misorientated from their parallel alignment thus suppressing the 
orbital overlap $S_{12}$. However, the contribution of the repulsion part to the 
interaction integral $H_{12}$ is much lower than that for $\theta=0^\circ$ and 
therefore, an enhancement of the splitting $(E_{1}-E_{2})$ due to suppression 
of $S_{12}$ is insignificant in comparison to that for the zero angle. As a result 
the size of the HOMO-LUMO gap is reduced only to $E_G$=1.65 eV that is much smaller 
than for $\theta=15^\circ$, but larger than for $\theta=0^\circ$. For AB stacking 
the HOMO-LUMO gap reaches its maximum because of the combination of the large 
orbital mixing and low orbital repulsion ($E_G$=2.01 eV).

Strong repulsion arising between stacked graphene flakes tend to increase the 
interlayer separation up to an equilibrium point when the attraction between the
layers cancel the effect of repulsion. To simulate this behavior we relaxed the 
system allowing the conformation changes only in the direction orthogonal to the 
graphene plane (full relaxation of the structure leads to sliding of the flakes 
away from each other). The equilibrium distance is found to deviate with the twist 
angle. For the AB stacked flakes for which the repulsion is found to be the 
weakest, the equilibrium distance is 4.21 \AA, while when the two lattices are 
superposed in AA stacking this distance approaches its maximum 4.99 \AA.
Therefore, as the increase in the twist angle between the flakes induces suppression 
of the interlayer repulsion, the equilibrium interlayer distance is reduced as 
shown in Fig.~\ref{fig:fig2} (b). The discrepancy of the magnitude of the 
interlayer distance calculated for small flakes (11.4$\times$9.8 \AA) 
from that of the standard value 3.4 \AA\ for the stacked graphene layers,
occurs due to the finite size effects and insufficient representation 
of the non-local dispersive interactions in the DFT techniques \cite{lebedev}.
In the case when the flakes are separated by the 
equilibrium distance depicted in Fig.~\ref{fig:fig2} (b),
the electronic properties resemble those 
for $AB$ stacking such that the HOMO-LUMO gap ($\sim$ 2.0 eV) remains almost unchanged for all 
values of the twist angles and so is the total energy of the system 
($E_{tot}$). The splitting of the (HOMO-(HOMO-1))
is insignificant (0.051 eV) thereby suggesting the efficient charge exchange between the 
two layers. Extrapolation of these results to the graphene layers of infinite size 
may lead to an important (and unique) situation which is described below. 

\begin{figure}
\includegraphics[scale=0.30]{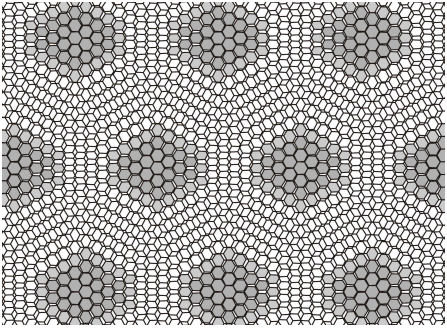}
\caption{\label{fig:fig4}
Image of the Moir\'e pattern for two graphene layers stacked with a rotation 
angle of 5$^{\circ}$. The shaded parts denote the areas where the two lattices 
are almost superposed.}
\end{figure}

The misorientation of stacked graphene layers observed experimentally 
\cite{hass,naturep,varchon,reina,warner} leads to the appearance of the Moir\'e 
pattern as shown in Fig.~\ref{fig:fig4}. In this pattern, some areas (spots 
shaded in Fig.~\ref{fig:fig4}) 
reveal the superposition of the graphene lattices (AA stacking), 
while in the inter-spot region the stacking similar to 
AB type is found to prevalent. The strong interlayer repulsion appearing 
in the areas of AA stacking tend to move the lattices apart, while in the 
rest of the system the prevalence of AB stacking tend to keep the graphene 
layers together. Therefore, in the areas of superposed lattice the repulsion 
would induce geometrical distortions in the form of 
{\it bumps} formed on the graphene surface. The diameter of 
each bump and its height (the bigger the spot size the weaker the elastic 
stress at the bump's boundaries) would be reduced with growing rotation angle. 
To confirm this statement we applied the geometry optimization 
(allowing the lattice relaxation in the direction orthogonal to the graphene plane)
to the stacked graphene flakes of size 17.5$\times$15.8 \AA\
for which the superposed and AB stacking areas are well defined for the rotation angle of 10$^{\circ}$.
Indeed, the formation of the bumps of height 
$\sim$0.20\AA\ was obtained on each flake. However, in the misoriented graphene layers 
we expect the bumps to be higher due to the smaller 
rotation angle and larger areas of inter-spot regions 
possessing the AB stacking where the repulsion will be weaker 
than in our model system. 
The suppression of the repulsion between the layers due to the lattice distortion 
makes the band structure uniform over the whole system.

In summary, two graphene layers stacked with a twist angle approaching $0^{\circ}$ 
display the electronic properties of a decoupled system because of the strong 
interlayer repulsion that arises between the interlayer located $\pi$ orbitals orthogonally 
positioned against each other after stacking. We determined that 
the misorientation in AA stacking observed in many experiments 
for non-exfoliated graphene \cite{varchon,naturep,hass,reina,warner} is meant to 
suppress that repulsion between the layers. The small twist angle leads to the
formation of the Moir\'e pattern with large areas (spots) where the graphene lattices 
are superposed. For the non-rigid systems the strong repulsion arising in the spot 
regions tend to move the two graphene lattices apart resulting in the formation of
bumps on the graphene surface. With increasing twist angle the areas with superposed lattices get smaller 
which minimizes the contribution of the repulsion to the geometry and the 
electronic properties of twisted graphene. This leads to an enhancement of the
electronic coupling between the layers which when becomes significant, changes the 
linear behavior of the bands to a parabolic one as evidenced in a bilayer graphene 
of AB stacking \cite{mccann,latil}. \\

The work was supported by the Canada Research Chairs program.


\begin{thebibliography}{99}

\bibitem{nov}
K.S. Novoselov, 
K.S. Novoselov, A.K. Geim, S.V. Morozov, D. Jiang, Y.
Zhang, S.V. Dubonos, I.V. Grigorieva, and A.A. Firosov,
Science {\bf 306}, 666 (2004).

\bibitem{review}
For a review on graphene, see, D.S.L. Abergel, V. Apalkov, J. Berashevich, 
K. Ziegler and T. Chakraborty, Advances in Physics {\bf 59}, 261 (2010).

\bibitem{varchon}
F. Varchon, P. Mallet, L. Magaud, and J.-Y. Veuillen, 
Phys. Rev. B {\bf 77}, 165415 (2008).

\bibitem{naturep}
G. Li, G. Li, A. Luican, J.M.B. Lopes dos Santos, A.H. Castro Neto, 
A. Reina, J. Kong, and E.Y. Andrei, 
Nature Phys. {\bf 6}, 109 (2009).

\bibitem{hass}
J. Hass, J. Hass, F. Varchon, J.E. Mill\'an-Otoya, M.S. Prinkle, 
N. Sharma, W.A. de Heer, C. Berger, P.N. First, L. Magaud, and E.H. Conrad, 
Phys. Rev. Lett. {\bf 100}, 125504 (2008).

\bibitem{reina}
A. Reina, A. Reina, X. Jia, J. Ho, D. Nezich, H. Son, V. Bulovic, 
M.S. Dresselhaus, and J. Kong,
Nano Lett. {\bf 9}, 30 (2009).

\bibitem{warner}
J.H. Warner,J.H. Warner, M.H. R\"ummeli, T. Gemming, B. B\"uchner, and G.A.D. Briggs, 
Nano Lett. {\bf 9}, 102 (2009).

\bibitem{mccann}
E. McCann, Phys. Rev. B {\bf 74}, 161403 (2006).

\bibitem{latil}
S. Latil, V. Meunier, and L. Henrard, Phys. Rev. B {\bf 76}, 201402 (2007).

\bibitem{shall}
S. Shallcross, S. Sharma, E. Kandelaki, and O.A. Pankratov, 
Phys. Rev. B {\bf 81}, 1 (2010);
Phys. Rev. Lett. {\bf 101}, 056803 (2008).

\bibitem{lopes}
J.M.B. Lopes dos Santos, N.M.R. Peres, and A.H. Castro Neto, 
Phys. Rev. Lett. {\bf 99}, 256802 (2007).

\bibitem{nbo}
Jaguar, version 7.5. Schr\"odinger. LLC: New York, NY, 2007. For details of 
NBO analysis see: http://www.chem.wisc.edu/$\sim$nbo5.

\bibitem{malard}
L.M. Malard, L.M. Malard, J. Nilsson, D.C. Elias, J.C. Brant, F. Plentz, 
E.S. Alves, A.H. Castro Neto, and M.A. Pimenta, 
Phys. Rev. B. {\bf 76}, 201401 (2007).

\bibitem{son-ritter}
Y.-W. Son, M. L. Cohen, and S. G. Louie, Phys. Rev. Lett. {\bf 97}, 
216803 (2006); K.A. Ritter and J.W. Lyding, Nat. Mater. {\bf 8}, 235 (2009).

\bibitem{barone}
V. Barone, O.Hod, and G.E. Scuseria, Nano Lett. {\bf 6}, 2748 (2006).

\bibitem{lebedev} 
I.V. Lebedeva, A.A. Knizhnik, A.M. Popov, Y.E. Lozovik, and B.V. Potapkin,
Phys. Chem. Chem. Phys.  {\bf 13}, 5687 (2011).

\end{thebibliography}
\end{document}